# Magnetic criticality-enhanced hybrid nanodiamond-thermometer under ambient conditions

Ning Wang[1,†], Gang-Qin Liu[1,†], Weng-Hang Leong[1,†], Hua-Ling Zeng[3,1,†], Xi Feng[1], Si-Hong Li[1], Florian Dolde[4], Helmut Fedder[4], Jörg Wrachtrup[4], Xiao-Dong Cui[5], Sen Yang[1,4], Quan Li[1,2,6,*] and Ren-Bao Liu[1,2,6,*]

1. Department of Physics, The Chinese University of Hong Kong, Shatin, New Territories, Hong Kong, China
2. The Chinese University of Hong Kong Shenzhen Research Institute, Shenzhen, China
3. ICQD, Hefei National Laboratory for Physical Science at the Microscale, and CAS Key Laboratory of Strongly-coupled Quantum Matter Physics, The University of Science and Technology of China, Hefei, China
4. Institute of Physics, Research Centre SCoPE and IQST, University of Stuttgart, 70569 Stuttgart, Germany
5. Department of Physics, The University of Hong Kong, Hong Kong, China
6. Centre for Quantum Coherence, The Chinese University of Hong Kong, Shatin, New Territories, Hong Kong, China

† These authors contribute equally

* Corresponding authors

**ABSTRACT:**

**Nitrogen vacancy (NV) centres in diamond are attractive as quantum sensors owing to their superb coherence under ambient conditions. However, the NV centre spin resonances are relatively insensitive to some important parameters such as temperature. Here we design and experimentally demonstrate a hybrid nano-thermometer composed of NV centres and a magnetic nanoparticle (MNP), in which the temperature sensitivity is enhanced by the critical magnetization of the MNP near the ferromagnetic-paramagnetic transition temperature. The temperature susceptibility of the NV center spin resonance reached 14 MHz/K, enhanced from the value without the MNP by two orders of magnitude. The sensitivity of a hybrid nano-thermometer composed of a $Cu_{1-x}Ni_x$ MNP and a nanodiamond was measured to be 11 mK/Hz$^{1/2}$ under ambient conditions. With such high-sensitivity, we monitored nanometer-scale temperature variation of 0.3 degree with a time resolution of 60 msec. This hybrid nano-thermometer**

**provides a novel approach to studying a broad range of thermal processes at nanoscales such as nano-plasmonics, sub-cellular heat-stimulated processes, thermodynamics of nanostructures, and thermal remanent magnetization of nanoparticles.**

**MAIN TEXT:**

Nanoscale temperature sensing is important for studying a broad range of phenomena in physics, biology, and chemistry, such as the temperature heterogeneities[1-3] in living cells, heat dissipation in nano circuits[4], nano-plasmonics, and nano-magnetism (like thermal remanent magnetism of nanoparticles). There have been a number of nano-scale temperature detection schemes[5, 6], such as scanning thermal microscopy (SThM)[7-9], SQUID based nano-thermometer[10], and fluorescence thermometers[11] based on rare-earth nanoparticles[12], nanogels[13], dyes[14] or proteins[15]. However, these techniques are limited by various factors, such as low sensitivity (rare-earth nanoparticles), contact-related artefacts (SThM), fluorescence instability (dyes), or the requirement of extreme working conditions (SQUID based nano-thermometer). Nitrogen-vacancy (NV) centres in diamond are a promising nano-sensor[16] owing to their atomic size and long spin coherence time[17, 18] under ambient conditions. The spin resonances of NV centres shift with temperature, providing a mechanism for atomic scale temperature sensors[19]. However, the temperature dependence of NV centre spin resonances, which results from the shift of the zero-field splitting $D$, is a small effect ($dD/dT \approx -74$ kHz/K)[19]. Therefore, even with single NV centres in ultra-high pure bulk diamond sample and advanced pulse control techniques, the sensitivity for temperature sensing is limited to a level of several mK/Hz$^{1/2}$ (Refs. 20-22).

Here we demonstrate the strategy of hybridization[23] to improve the sensitivity of diamond nano-thermometers by converting the temperature variation to a magnetic field change. NV centres have been demonstrated to be ultra-sensitive to external magnetic field[24-28]. The magnetization of a magnetic nanoparticle (MNP) can be used to monitor its local temperature, and this mechanism becomes ultra-sensitive when temperature is close to the magnetic phase transition point (i.e., critical point) of the MNP. The sensitivity of a hybrid nano-thermometer composed of a $Cu_{1-x}Ni_x$ MNP and a nanodiamond under ambient was measured to be 11 mK/Hz$^{1/2}$ under ambient conditions. The temperature range in which the hybrid sensor has high sensitivity can be setting different critical temperatures of the materials (e.g., via tuning the chemical composition of the $Cu_{1-x}Ni_x$ MNP). This hybrid nano-thermometer provides a novel approach to studying a broad range of thermal processes at nanoscales such as nano-plasmonics, sub-cellular heat-stimulated processes, thermodynamics of nanostructures, and thermal remanent magnetization of nanoparticles.

## Results

**Scheme of the hybrid sensor and theoretical sensitivity estimation**

The hybrid nano-thermometer is composed of a fluorescent nanodiamond (FND) and an MNP, as shown in Fig. 1a. The ground state of the NV centre is a spin triplet (S = 1) with a zero-field splitting D ≈ 2.87 GHz between the $m_s = 0$ state and the $m_s = \pm 1$ states. The spin transitions of NV centres in the FND are shifted by the magnetic field from the MNP, $B_{MNP}(T)$. The Hamiltonian of an NV centre can be written[16] as

$$H_{\text{eff}} = D(T)S_Z^2 + E(S_X^2 - S_Y^2) - \gamma \mathbf{S} \cdot \mathbf{B}_{\text{MNP}}(T), \quad (1)$$

where $E$ is the lattice strain effect and $\gamma$ = 28 MHz/mT is the electron gyromagnetic ratio. The field from the MNP $\mathbf{B}_{MNP}(T)$ is determined by its magnetization $\mathbf{M}(T)$, which depends sensitively on the temperature when the temperature is below and close to the ferromagnetic-paramagnetic transition point (Fig. 1b). Therefore the spin resonance of the NV centres in the FND presents abrupt dependence on temperature near the transition point of the MNP. Previously, the temperature shift of zero filed splitting d$D$/d$T$≈74 kHz/K is measured to extract the temperature change[19-22]. Here the transition frequencies $\omega_\pm$ (for $|0\rangle \leftrightarrow |\pm 1\rangle$) of the NV centre spins are measured to extract the magnetization properties of the adjacent MNP (Fig. 1c), which serves as a transducer and amplifier of the local temperature variation. The critical behavior of the MNP enhances the temperature sensitivity of the NV centres.

The sensitivity of the hybrid sensor is determined by the magnetization of MNP, the number and coherence of NV centres, and the relative position and orientation between the two nanoparticles. FNDs with size ~100 nm containing 500 NV centres were used in experiments. As for MNPs, Copper-nickel alloy nanoparticles were chosen for their chemical stability under ambient conditions and magnetization stability and repeatability due to large anisotropic energy. To estimate the sensitivity of the hybrid nano-sensor, we consider a spherical $Cu_{0.30}Ni_{0.70}$ MNP of diameter 200 nm located 50 nm away from an FND. Taking into account the strain distribution in the FND, the spatial and orientation distributions of the NV centres, and the magnetic field gradient from the MNP, we estimate the optimal sensitivity to be about 3 mK/Hz$^{1/2}$, as plotted by black dots in Fig. 1d. This represents an improvement by three orders of magnitude from the sensitivity of a bare FND sensor (the previously record of similar FND sensors is 200 mK/Hz$^{1/2}$ as reported in Ref. 20).

The sensitivity enhancement of the magnetic criticality-enhanced sensor is at the cost of working range (the optimal range being ~ 20 °C below the Curie temperature). This disadvantage can be compensated by choosing MNPs of different Curie temperatures. Taking $Cu_{1-x}Ni_x$ MNPs as an example, their Curie temperature is tuneable from near 0 K to 637 K with the nickel composition $x$ from 45 to 100%[31]. Together with the very broad working temperature of NV centres in diamond[32,33], the hybrid sensor can be designed to fulfil a wide range of applications. In Fig. 1d, we present the composition (and thus temperature) dependence of the sensitivity of the $Cu_{1-x}Ni_x$ MNP based hybrid sensor. The optimized temperature sensitivity is better than 10 mK for the whole range (0 K to 637 K), which has considered the contrast change of optically detected magnetic resonance (ODMR) with temperature[32].

**Proof-of-the-principle experiment – Gd particle and bulk diamond**

As a proof-of-the-principle demonstration of the magnetic criticality-enhanced quantum sensing, we tested a hybrid sensor composed of a gadolinium (Gd) particle (with size ~2 mm and mass ≈30 mg) and a single NV centre in a bulk diamond (3 mm

x 3 mm x 1 mm). The distance between the NV center and the Gd particle is about 2 mm. Such a distance is chosen to approximately simulate the configuration of an MNP coupled to an FND since the field from a magnetic particle, in the dipole approximation, is proportional to the total volume and inversely proportional to the cubic distance. Gd has the ferromagnetic-paramagnetic transition near room temperature ($T_C \approx 19\,°C$)[34]. Continuous-wave ODMR measurement was carried out in a temperature range between 11 °C and 37 °C under a uniform external magnetic field ($B \approx 100$ Gauss). The temperature was controlled by an incubator (INSTEC) and independently monitored by a Pt thermocouple.

Figure 2a presents the ODMR spectra of the hybrid sensor at various temperatures. The frequency shifts of the $m_s = 0 \leftrightarrow \pm 1$ transitions (Fig. 2b) are enhanced near the critical point of Gd, with $d\omega/dT$ reaching its maximum value $\approx 14$ MHz/K at 19 °C. The induced magnetic moment can be deduced from the ODMR spectrum, which is consistent with the magnetic moment of the same Gd particle measured by vibrating sample magnetometer (VSM, Quantum Design). For comparison, the ODMR spectral shift of the same NV centre (with Gd removed) is shown in Fig. 2c, which presents only a weak dependence on temperature, $dD/dT = -71 \pm 2$ kHz/K (Fig. 2d), consistent with previous reports[19]. The magnetic criticality of the Gd particle induces an enhancement of spectral susceptibility to temperature by a factor of 200. The magnetically enhanced temperature sensitivity of the NV centre is evaluated with Eq. (3) in Method using the experimentally measured $d\omega/dT$, the spin resonance width, the contrast, and the photon counts. We realized a sensitivity of $45$ mK Hz$^{-1/2}$. While this sensitivity was limited by the weak photon counts ($8\times10^4$ sec$^{-1}$ per NV center) in our system, it is already 200 times better than the bare NV centre in the same setup.

**Sensitivity enhancement by ensemble NV centres – Gd particle and FND**

To achieve a better temperature sensitivity of the hybrid sensor, we replaced the single-NV bulk diamond with an FND containing ensemble NV centers (from Admas, claimed of about 500 NV centres per FND) to enhance the photon counts of the hybrid sensor. The FND was placed about 2 mm away from the Gd particle. The distance between the FND and the Gd particle was chosen comparable to the size of the Gd particle so as to simulate the field strength from an MNP. To show that the dependence of the sensitivity on temperature, we measured the sensitivity at different temperatures with a photon count rate of $6\times10^6$ sec$^{-1}$. The resonant frequency $\omega_-$ as a function of temperature is shown as an inset in Fig. 3a (with a sharp change near $T_C$). The sensitivity, as determined by equation (3) in Method using experimentally measured photon counts, spin resonance width, contrast and d$\omega$/d$T$, is shown in Fig. 3a (with a sharp change near $T_C$).

For optical detection sensing, a fundamental limitation to sensitivity is the shot noise of photon counts[27]. To check whether the sensitivity in our system is limited by photon shot noise, we measured the temperature for various photon counts (by using various laser powers) and for various data acquisition time. The three-point CW ODMR method is adapted from Refs. 20 & 35 for fast temperature determination (see Method). The temperature range was chosen near the optimal value ($T\approx19$℃, close to $T_C$ of Gd). Figure 3b shows the standard deviation of the measured temperature $\delta T$ as a function of the data acquisition time (see Method). With long enough integration time for all the different photon counts, the minimum measurable temperature variation $\delta T$ approaches to a lower bound ~ 10 mK, the temperature stability of the incubator. The linear dependence of $\delta T$ on the inverse square root of integration time gives the photon shot noise limited sensitivity. As expected, higher photon counts lead to higher sensitivity. The sensitivities derived from the dependence of $\delta T$ on the data acquisition time agrees well with the measured sensitivity using equation (3) in Method.

**Nano-thermometer – CuNi MNP and FND**

Finally, we demonstrate that the criticality-enhanced sensing works at nanoscale. Since gadolinium nanoparticles are easily oxidized under ambient conditions, we use copper-nickel (CuNi) MNPs instead of Gd nanoparticles in the hybrid nano-sensor. As shown in Fig. 4a, the pattern matching between the confocal image and the transmission electron microscopy (TEM) imaging identifies an FND located in the proximity of a CuNi MNP. The temperature change of the CuNi MNP is induced by in-situ laser heating of the carbon film on the TEM grid (the direct heating of the MNP and the FND by laser and microwave is negligible). The temperature at the FND was calibrated by the zero-field splitting $D$, which is assumed identical to the temperature at the MNP since the distance between the two particles is less than 100 nm and the laser spot is ~ 300 nm. No external magnetic field was employed in the nano-thermometer experiments.

The temperature response of the nano-thermometer is determined by measuring the resonant transitions frequencies $\omega_{\pm}$ of NV centers under the magnetic field from the MNP at different temperatures. The gradient of the magnetic field from the CuNi MNP and the spatial distribution of NV centres in the FND cause extra broadening in the ODMR spectra, as shown in Fig. 4b upper panel. Nevertheless, the ODMR dips and the Zeeman splitting are well discernible even in the presence of the gradient-induced broadening. For all laser powers (hence temperatures), the ODMR spectra are well consistent with the theoretical simulation that includes the Zeeman splitting, the MNP gradient-induced broadening and the strain distribution of ensemble NV centres in the FND.

The frequency shift induced by the magnetization of the MNP is reversible when the temperature was scanned back and forth between below and above the critical temperature of the material, as in Fig. 4c. The reversibility is ascribed to the large anisotropy energy of CuNi nanoparticles. Meanwhile, this hybrid sensor is very robust and stable. In measurements over more than one month, its temperature response showed negligible change. From the frequencies resonance, the magnetization of the CuNi MNP at different temperatures can be determined by numerical simulation, which agrees well with the critical behavior of CuNi (Fig. 4d). The Curie temperature is determined to be about 340 K, consistent with the composition of the MNP $Cu_{0.26}Ni_{0.74}$ measured by the energy-dispersive X-ray spectroscopy (EDX) in TEM[31]. The consistency verifies that the heating within the 300 nm laser spot is uniform and the temperatures of the MNP and the FND are close.

Then we measured the sensitivity of the hybrid nano-sensor. In the nano-sensor, due to the magnetic field gradient and NV centre distribution in the FND, the line shape changes (both line width and contrast) much more significant than the frequency shift with temperature, especially for temperatures close to the critical point. As illustrated in Fig. 4b, in the 1 ℃ temperature change close to $T_C$ (67 ℃), the ODMR line shape changes significantly. From the difference between the two spectra, we find that the most sensitive frequency is close to the center of the dip (Fig.4b lower panel). Thus we use the contrast change d$S$/d$T$ for the microwave frequency fixed at the dip position to measure the sensitivity. According to equation (2) in Method, a sensitivity of 11 mK/Hz$^{1/2}$ was realized at the most sensitive point (the temperature-induced variation of the normalized signal d$S$/d$T$=0.025 /K at $\omega = 2866$ MHz, with photon counts L= 12 Mcps). This sensitivity is close the the optimal temperature sensitivity of 3 mK/Hz$^{1/2}$ (Fig. 1d).

We tested the real-time temperature monitoring using the hybrid nano-sensor. With the modified three-point method (choosing two of the most sensitive frequencies, at the dip of the ODMR spectrum, one far from the resonance, as reference to cancel laser fluctuation), we carried out the real time measurement at 63℃, the shot-noise limited sensitivity for three-point method could be 23(2) mK/Hz$^{1/2}$, as shown in Fig. 5a. Then we generated a periodic temperature alternation between two values (63 ±0.75℃) near the Curie temperature of the CuNi MNP by modulating the intensity of the excitation laser (the temperature is calibrated by the D shift as function of laser power). With the tree-point method, we can monitor the temperature variation with acquisition time 60 ms per data point, as shown in Fig. 5b. Of the two temperature states, the signal can reach at the same level of each period, indicating that the reversibility and stability of this hybrid sensor are quite well.

## Discussion

The measured sensitivity of an NV based thermometer is determined by the sample quality (number and coherence of NV centres), the detection protocols (CW or pulsed measurement), and the fluorescence collection efficiency of the experimental setup. A setup- and protocol-independent parameter is $d\omega/dT$. Two orders enhancement of $d\omega/dT$ is demonstrated for all the three hybrid thermometers, which is well consistent with the theoretical estimation. The universal enhancement of magnetic criticality leads to a temperature sensitivity as high as several mK/Hz$^{1/2}$, even with NV centres of poor coherence and only the continuous continuous-wave ODMR technique. By comparing with the theoretical calculation, we conclude that the temperature sensitivity of the hybrid sensor can be improved by optimizing the fluorescence collection efficiency of the confocal microscopy, the microwave excitation efficiency of the antenna, and the relative orientation between NV and the magnetization axis of MNP.

The temperature sensitivity can be further improved by using NV centre spins of longer coherence times and employing pulse control protocols for high-sensitivity d.c. magnetic field sensing[26]. Figure 6a shows the configuration of a hybrid sensor composed of a 200 nm Cu$_x$Ni$_{1-x}$ MNP on a diamond nano-pillar and a single NV centre (photon counts $L$=1.7 Mps and coherence time $T_2^* \approx 10~\mu s$). For a CuNi particle, the sensitivity is plotted in red dots of Fig. 6b with typical Ramsey sequence (shown inset of Fig. 6b). A sensitivity of $1~\mu K/Hz^{1/2}$ can be achieved near Curie temperature. If a single NV centre in isotopically purified diamond (99.99% $^{12}$C) is used, with the ultra-long coherence time ($T_2^* \approx 250~\mu s$), the sensitivity could be 0.3 $\mu K/$ Hz$^{-1/2}$, as shown as navy dots in Fig. 6b.

With the superb sensitivity and designable working temperature range, the hybrid thermometer can enable in situ and real-time observation of heat dissipation at nanoscale. Systems of interest include operating microelectronic devices with non-uniform heat dissipation[4], nano-plasmonics, subcellular thermal processes, and chemical reaction at nanoscale[36]. The FND/CuNi MNP nano-thermometer can work under zero field, which would allow the setup be miniaturized and the sensor be applied to liquid environments where the rotation is inevitable. Furthermore, ND based sensors are biocompatible and robust against systematic errors such as fluorescence fluctuation in the complicated intracellular environment. These unique merits enable directly monitoring of cell metabolism, and make it possible to clarify the controversies on temperature imaging in single living cells[1-3,37].

# Methods

**Sensitivity – theoretical estimation and experimental measurements**

Magnetizations of $Cu_{0.3}Ni_{0.7}$ alloy NMP are calculated with a mean-field theory. Transitions frequencies of the NV centre spins are obtained by diagonalizing the effective Hamiltonian in equation (1). For ensemble NV centres in an FND, continuous wave ODMR is utilized for thermal sensing. Thus sensitivity[28] becomes

$$\eta_{CW} \approx \frac{1}{\sqrt{L}} \left|\frac{dS(\omega)}{dT}\right|^{-1}_{max}, \qquad (2)$$

where $S(\omega)$ is the normalized ODMR spectrum and $L$ is the photon count rates. If the ODMR spectrum is Lorentz shape. The sensitivity can be optimized by choosing the frequency at the half-height of the resonance. The result is

$$\eta_{CW} \approx \frac{4}{3\sqrt{3}} \frac{\Delta\omega}{C\sqrt{L}(d\omega/dT)}, \qquad (3)$$

where $\Delta\omega$ is the linewidth and $C$ is the contrast of the resonance.

To test whether the sensitivity of our system is shot noise limited, we kept the system at a stable temperature and different integration times were used to carry out three-point ODMR measurements. One point was chosen as the reference frequency to normalize the laser fluctuation for long term measurement. The sensitivity evaluated from three-point method is larger by $\sqrt{1.5}$ than the sensitivities evaluated from Equation (2) and (3). The standard deviation of temperature $\delta T$ as a function of the square root of integration time $\sqrt{\Delta t}$ determines the temperature sensitivity, $\eta = \delta T \cdot \sqrt{\Delta t}$.

**Sample preparation**

The bulk diamond (from Element Six) is a high-purity type-IIa sample with a natural 1.1% $^{13}$C abundance. The FNDs (from Adamas Nanotechnologies) are of type Ib, prepared by high-pressure high-temperature synthesis. Each FND contains about 500 NV centres as claimed by Adamas and confirmed by the high fluorescence rate.

The CuNi MNPs were prepared by ball milling of micron size $Cu_{0.30}Ni_{0.70}$ alloy (from Sigma-Aldrich). The FNDs and the CuNi MNPs were dispersed in ethanol separately. Then the two solutions were subsequently dropped on a TEM gold grid. The size, composition and relative positions of the two types of nanoparticle were characterized by TEM techniques. The featured patterns of carbon film on the TEM grid were used to match the confocal and TEM images.

**Experimental setup**

We used a home-built confocal microscope for imaging and ODMR measurement. NV centre spins are optically pumped by a 532 nm laser, manipulated by resonant microwave fields applied through a 25 μm diameter gold wire, and detected via spin-state-dependent fluorescence measurements. The power of the microwave is adjusted to maximize the ODMR contrast but without extra heating effect and is kept unchanged through the experiments. For FNDs, the width of ODMR spectra is mainly determined by the inhomogeneous broadening of ensemble NV centres (which is much greater than the power broadening induced by the laser excitation). For the CuNi-FND hybrid sensor, the nanoscale magnetic field gradient causes extra broadening to the ODMR spectra.

An incubator from INSTEC was used in the Gd-related experiments to control the temperature with the stability of sub 10 mK. A Pt thermocouple was placed near the Gd particle to monitor the local temperature. The heating and cooling rate of the incubator is 0.5 ºC/min, and ODMR measurements were carried out after the sample reached its thermal equilibrium state (in about 20 minutes). In the real time temperature monitor experiments (Fig. 3c), the heater of the incubator generates a periodical temperature variation (with square-wave of 0.2 ºC amplitude and 400 sec period), and induces a smooth temperature oscillation at the sample position (with 0.1 ºC amplitude and 400 sec period, due to the thermal gradient from the heater to sample).

In the experiments on the CuNi MNP-FND sensor, the MNPs were in-situ heated by light absorption of the carbon film on the TEM grid. The local temperature of the hybrid sensor was tuned by laser intensity, which was controlled by the acoustic optical modulator (AOM) before the fiber entrance of the microscope. The liquid environment of the hybrid sensor (oil immersed) stabilizes the laser heating effect. The measured thermal stability of laser heating on TEM grid is better than 0.5 ºC, in the temperature range from room temperature (25 ºC) to about 70 ºC (which is calibrated from the zero splitting $D$ of the FND).

**Acknowledgements** This work was supported by the National Basic Research Program of China (973 Program) under Grant No. 2014CB921402, Hong Kong Research Grants Council - Collaborative CUHK4/CRF/12G, and The Chinese University of Hong Kong Vice Chancellor's One-off Discretionary Fund. H.L.Z. acknowledges the support from China Government Youth 1000-Plan Talent Program and National Natural Science Foundation of China under Grant No. 11674295.

**Author Contributions** R.B.L. & Q.L. conceived the idea, designed the project, and supervised the research, W.H.L. carried out the theoretical studies and numerical simulation, N.W., G.Q.L., H.L.Z., Q.L. & R.B.L. designed the systems and measurements, G.Q.L., N.W., H.L.Z., S.H.L., F.D., S.Y., X.D.C. & J.W. set up the ODMR system, N.W. & G.Q.L. carried out the experiments, N.W. & F.X. processed the samples, F.X. characterized MNPs and FNDs on TEM grids, N.W., G.Q.L., H.L.Z., W.H.L., X.F., Q.L. & R.B.L. analysed the data, N.W. wrote the paper, and all authors commented on the manuscript.

**Author Information** The authors declare no competing financial interests. Correspondence and requests for materials should be addressed to R.B.L. (rbliu@cuhk.edu.hk) or Q. L. (liquan@phy.cuhk.edu.hk)

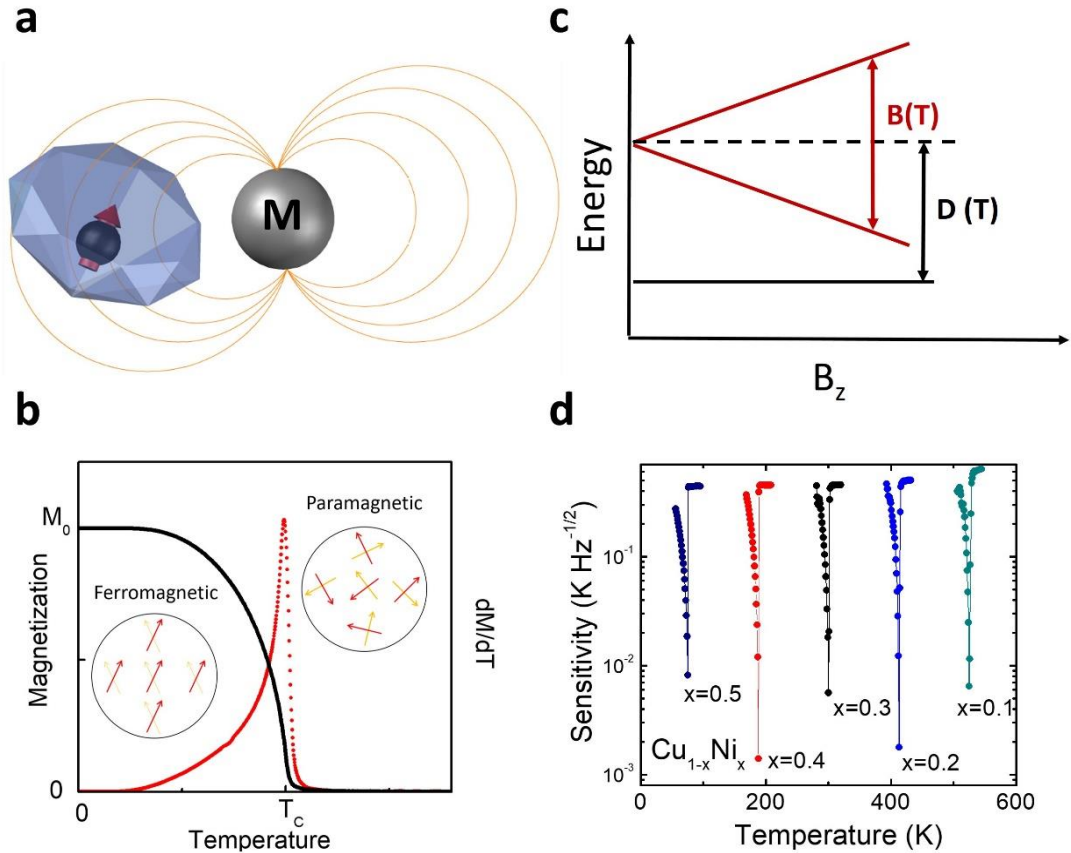

**Figure 1 | Principle of magnetic criticality enhanced hybrid nano-thermometers.**
**a,** Schematic of a hybrid nano-sensor, composed of a fluorescent nanodiamond (FND) and a magnetic nanoparticle (MNP). **b,** The magnetic moment $M$ (black solid line) of the MNP decreases dramatically with temperature increasing near the ferromagnetic-paramagnetic transition, with the temperature susceptibility $dM/dT$ (red dotted line) peaking at the critical temperature $T_C$. The transition frequencies of the NV centre spins in the FND depend sensitively on the magnetic moment of the MNP and hence the temperature. **c,** Fine structure of the NV centre ground state as a function of an axial magnetic field. The zero-field splitting ($D$) varies with temperature with a rate $dD/dT \sim$ -74 kHz/K. In the hybrid nano-sensor, the magnetic field from of the MNP $B(T)$ results in a much more sensitive dependence of the transition frequencies on the temperature. **d,** Theoretical sensitivity estimation of the hybrid sensor of an FND and a $Cu_{1-x}Ni_x$ MNP with different chemical composition ($x$). A spherical CuNi MNP with diameter of 200 nm is assumed. The FND is assumed of size ~ 100 nm with 500 NV centres and located 50 nm away from the MNP.

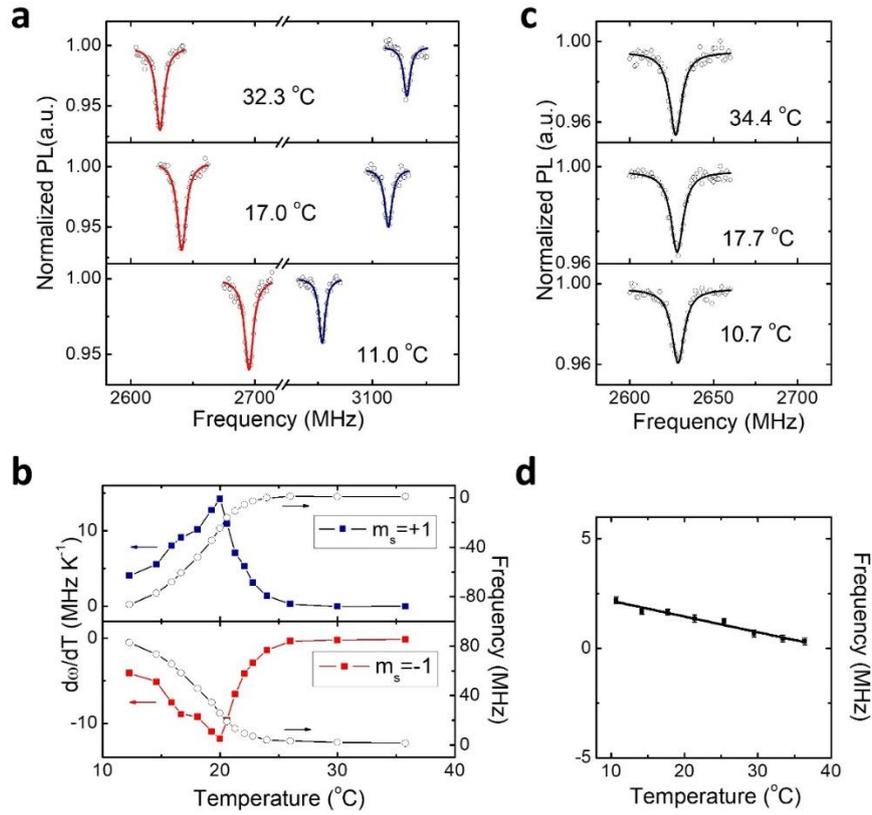

**Figure 2 | Proof-of-the-principle demonstration of magnetic criticality enhanced temperature sensing.** The ODMR spectra of an NV centre in a bulk diamond near a Gd particle are compared to the spectra of the same NV centre without the Gd particle. An external magnetic field $B = 100$ Gauss is applied. The temperature is varied near the critical point of Gd. **a**, With the criticality enhancement, the hybrid sensor shows a large spectral shift with temperature. Left/right resonances (red/blue lines) correspond to the $m_S = 0 \leftrightarrow m_S = -1/+1$ transitions. **b,** The ODMR frequencies (black solid lines) of the hybrid sensor versus the temperature. The temperature susceptibility d$\omega$/d$T$ (square dots) of the hybrid sensor reaches ±14 MHz/K at the critical point $T_C$=19 ºC, which is 200 times larger than that of the bare NV sensor. **c**, A bare NV centre shows only a slight spectral shift of the $m_S = 0 \leftrightarrow -1$ transition under the same temperature change as in **a.** In (**a**) and (**c**), the dots are experimental data and the lines are the Lorentzian fitting. **d,** The ODMR frequencies of the same NV sensor as a function of the temperature, which has a slope of -74 kHz $K^{-1}$.

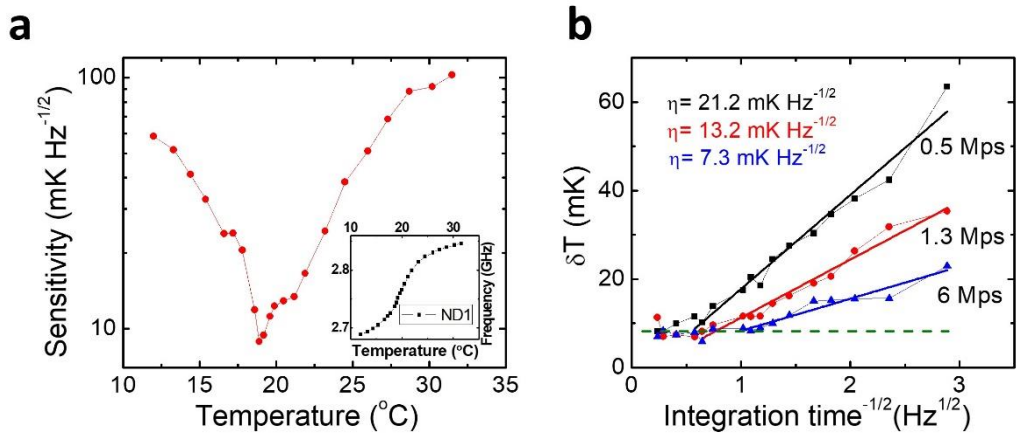

**Figure 3 | Sensitivity of a magnetic criticality enhanced Gd-FND temperature sensor. a,** The sensitivity of the hybrid sensor at various temperature. The black dots are sensitivity determined by Equation (3) using the experimentally measured photon counts, contrast, $d\omega/dT$ and spin resonance width $\Delta\omega$. Inset: the frequency shift of the Gd-FND sensor as a function of temperature. The laser power is chosen such that the photon count is $\sim 6\times 10^6$ sec$^{-1}$. The optimal sensitivity ~7.3 mk is achieved at $T_C$ = 19 ºC. **b**, The standard deviation of temperature measurement as a function of data acquisition time for various photon counts (due to different laser powers) using 3pts method. The slopes of the curves give the shot-noise limited sensitivity of the hybrid sensor. The green dash line corresponds to the temperature stability of the incubator.

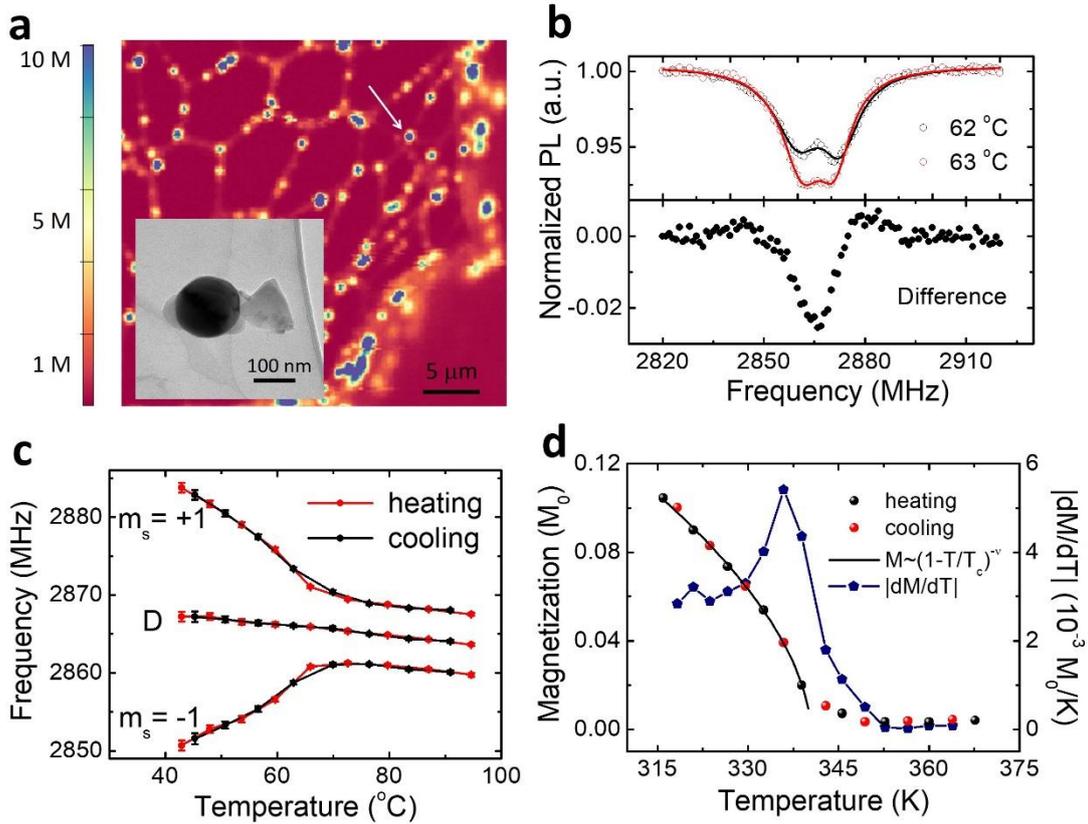

**Figure 4 | Magnetic criticality enhanced hybrid nano-thermometer.** The nano-thermometer is composed of an FND (with about 500 NV centres inside) and a nearby $Cu_{1-x}Ni_x$ MNP. **a,** Confocal image of the hybrid sensor (inset, TEM image. The black solid one is a $Cu_{1-x}Ni_x$ MNP and the transparent one is an FND). **b**, ODMR spectra of the hybrid sensor for various temperatures. No external magnetic field was applied. The splitting is caused by the magnetic field from the CuNi MNP. Lower panel is the signal difference of the two ODMR spectra in the upper panel. **c,** ODMR resonant frequencies of the $m = 0 \leftrightarrow m = \pm 1$ transitions as functions of temperature. The zero-field splitting $D$ is obtained by averaging the two frequencies. Both heating (red dots) and cooling (black dots) processes are measured. **d,** The spontaneous magnetization of the CuNi MNP extracted from the ODMR spectrum. The red and black dots are measured in the heating and cooling processes, respectively. The blue curve shows the temperature susceptibility of the magnetization. The transition temperature of this nanoparticle (depending on the Ni composition *x*) is about 340 K, which is consistent with its chemical composition ($Cu_{0.25}Ni_{0.75}$) measured by EDX in the TEM.

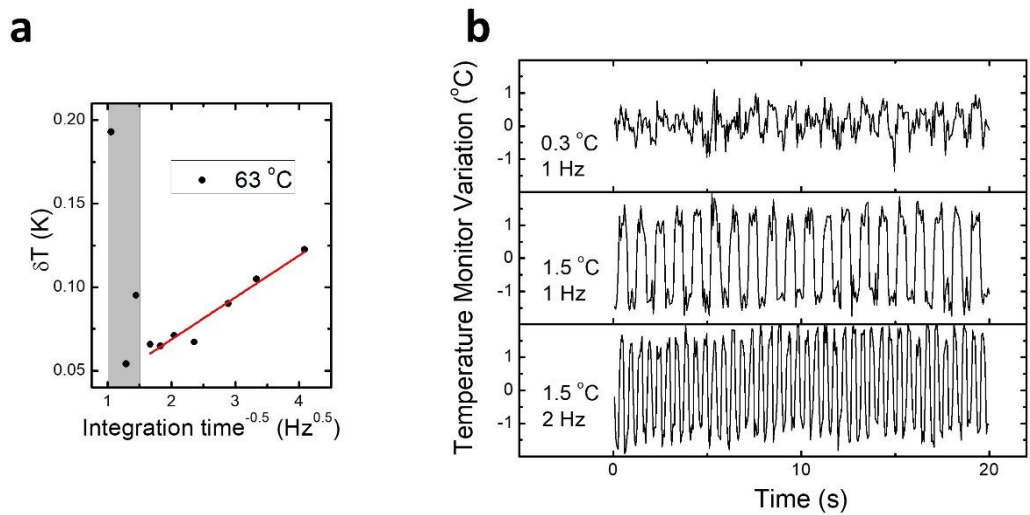

**Figure 5 | Shot-noise limited sensitivity and real-time temperature tracking.** The sensor is the same one as used in Fig. 4. **a**, The standard deviation of temperature measurement as a function of data acquisition time using 3pts method. The slope of the curve gives the shot-noise limited sensitivity of the hybrid sensor. **b**, The hybrid sensor was used to monitor a fast temperature variation at nanoscale. The local temperature was controlled by in-situ laser heating, which has tunable amplitude and modulation frequency (by an acoustic optical modulator, AOM). The temperature (about 63 ℃) is close to the Curie temperature of the CuNi MNP and therefore a high sensitivity is achieved. Acquisition time for each data point is 60 ms. The hybrid sensor shows perfect reversibility during the whole measurement.

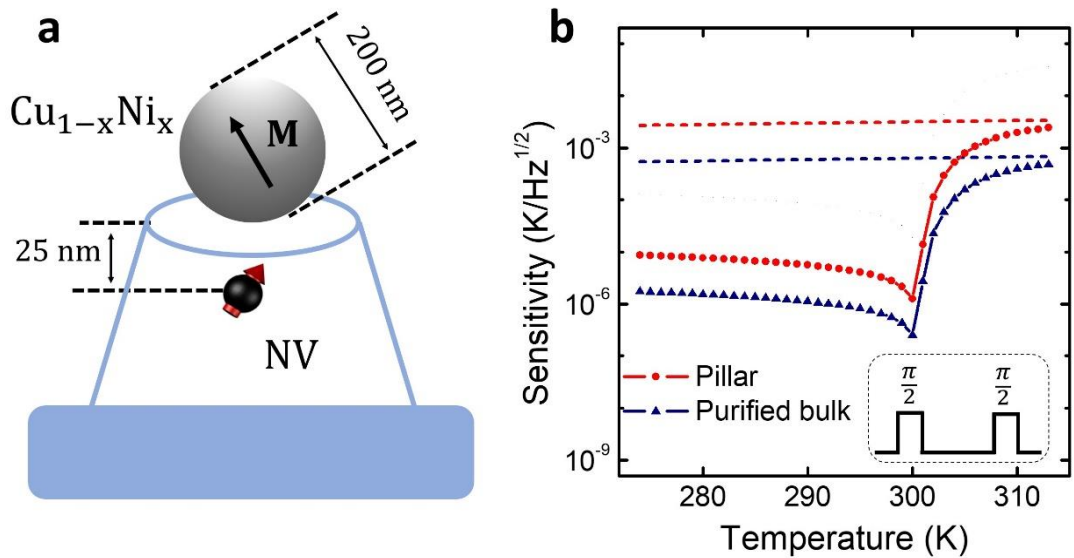

**Figure 6 | Theoretical sensitivity of a hybrid nano-thermometer composed of an MNP and a single NV center.** With the nanostructure enhanced fluorescence collection efficiency and long coherence time of single NV centers, this hybrid sensor gives much high temperature sensitivity. **a,** Schematic of the hybrid sensor that contains an MNP on the top of a diamond nano-pillar with a single NV centre 25 nm below the surface. **b,** Sensitivity versus temperature for two configurations of the hybrid sensor. The red solid circle is constructed by a diamond nanopillar with natural abundance $^{13}$C (1.1%), and the blue triangle represents the results from an isotopically engineered one, which has long coherence time. With pulse control technique (shown in inset), the long coherence time can be fully exploited. The dash lines are the sensitivities of the bare NV-based temperature sensors, red for the NV center in the natural abundance $^{13}$C diamond pillar and blue for the isotopically engineered one.